\UseRawInputEncoding
\documentclass[12pt]{iopart}

\usepackage{graphicx}
\usepackage{iopams}
\usepackage{subfigure}
\usepackage{rotating}
\usepackage{color}

\begin{document}

\title{Formation of quasi-single helicity state from a paramagnetic pinch in KTX regime}

\author{Bing Luo$^{1}$, Ping Zhu$^{2,3}$, Wentan Yan$^{4}$, Hong Li$^{4}$ and Wandong Liu$^{4}$}
\address{$^1$ Hunan Province Key Laboratory Integration and Optical Manufacturing Technology, College of Mathematics and Physics, Hunan University of Arts and Science, Changde, Hunan 415000, China}
\address{$^2$ International Joint Research Laboratory of Magnetic Confinement Fusion and Plasma Physics, State Key Laboratory of Advanced Electromagnetic Engineering and Technology, School of Electrical and Electronic Engineering, Huazhong University of Science and Technology, Wuhan, Hubei 430074, China}
\address{$^3$ Department of Nuclear Engineering and Engineering Physics,University of Wisconsin-Madison, Madison, Wisconsin 53706, USA}
\address{$^4$ School of Nuclear Science and Technology, University of Science and Technology of China, Hefei, Anhui 230026, China}

\ead{zhup@hust.edu.cn}
\vspace{10pt}
\begin{indented}
\item[] August 2024
\end{indented}

\begin{abstract}
The formation of quasi-single helicity (QSH) state from a paramagnetic pinch in the KTX-RFP regime has been observed in recent NIMROD simulations. The quasi-single helicity state has a dominant helical component of the magnetic field that is known to improve the RFP confinement. For the initial paramagnetic pinch, linear calculations indicate that the tearing mode growth rate decreases with the plasma $\beta$. The initial QSH state arises from the dominant linear instability of the initial force-free paramagnetic pinch. The plasma's self-organization towards the second QSH state after the relaxation of the initial QSH state is found to depend on $\beta$. Specifically, when $\beta<4\%$, the plasma relaxes to an MH state; when $4\% \leq \beta \leq 8\%$, the plasma first transitions from a double axis (DAx) to a single helical axis (SHAx) state, and eventually return to the DAx state. The existence of such an optimal $\beta$ regime that is beneficial to the formation and maintenance of the QSH state, suggests an experimental scheme for the QSH formation based on $\beta$ tuning and control.
\end{abstract}

%

%
%
%

\section{Introduction}

~~~~The reversed-field pinch (RFP) [1] is a type of magnetic confinement of plasma that uses the pinch effect due to a current flowing in a toroidal magnetic field. The toroidal ($B_\phi$) and poloidal ($B_\theta$) components of the magnetic field are mainly created by the plasma itself and are of of similar strength. The RFP plasma displays various states due to self-organization. The safety factor of RFP $q(r) < 1$, which peaks near or on the magnetic axis and gradually decreases to zero at or near the plasma edge, leading to the presence of many unstable resonant tearing modes with the poloidal mode number $m=1$ and various toroidal mode number $n_s$ at different radii [2]. In typical RFP discharges, the $m=1$ Fourier components of the magnetic field perturbation have comparable amplitudes, resulting in the multi-helicity (MH) state. After the corresponding islands overlap in radial width, the magnetic stochasticity ensues and degrades the energy confinement [3].

As the plasma current increases, the innermost resonant mode can grow to a large amplitude, creating a state known as quasi-single helicity (QSH) [4,5]. This state, observed in various RFP devices, is characterized by the presence of a dominant $m=1$ mode with $n > 3R_0/2a$, leading to the formation of high-temperature 3D helical flux surfaces in the plasma core, similar to stellarator configurations [4,6-11]. The QSH state can be of double axis (DAx) or single helical axis (SHAx) type [10,12,13]. The occurrence of the QSH state reduces magnetic chaos, and thus improves the RFP confinement, especially for the SHAx state with a relatively wide hot helical core and strong transport barrier [14-16]. Experimental statistics from RFX-mod show that the QSH properties are related to the plasma current or Lundquist number $S$, with higher plasma current leading to longer persistence (up to about $90\%$) of the QSH state [16]. The Virtual Shell (VS) has beneficial effects on plasma confinement, the Clean Mode Control (CMC) has allowed for the increased plasma current, and the edge helical magnetic perturbations (MPs) from external active coils have been used to excite the QSH with specific dominant modes in RFX-mod [17-20].

Earlier, E. J. Caramana utilized the single-fluid MHD model to study the single helicity state in RFP [21]. The characteristics of single helicity and multiple helicity ohmic states in RFP were investigated with both analytical and numerical methods [22]. Simulations using the 3D resistive magnetohydrodynamics SpeCyl code suggest that the transition from MH to SH states is governed by the Hartmann number $H ={a V_{A}}/ {\sqrt{{\eta_{m}}{\nu}}}$, where $\eta_{m}=\eta/\mu_0$ represents the magnetic diffusivity, $\eta$ is the resistivity, and $\nu$ the kinematic viscosity [23-27]. It was observed in simulations that in the stationary SH regime, charge separation exhibits a significant dipolar helical component [28]. The use of helical boundary conditions generated by active coils to excite and control the specific dominant modes of the QSH state in RFP has also been demonstrated in simulations [19,20,29,30]. In the KTX configuration and regimes, the 3D characteristics and dynamo effect were investigated based on the VMEC equilibrium, and the evolution of self-organized state was studied using the SPEC code [31-33].

On the other hand, previous NIMROD simulations also suggest that the plasma $\beta$ may be another key parameter that can significantly influence the emergence and duration of the QSH state [34]. However, these simulations start directly from the 2D RFP equilibrium [22,24,35], which is not necessarily in the ohmic steady state. By contrast, the paramagnetic pinch is the natural final configuration of a driven plasma in the absence of any instabilities, and it satisfies the ohmic steady state [36,37]. It is thus interesting for us to explore in this work if or how the QSH state may still form from the initial paramagnetic pinch, and whether $\beta$ remains to be a key controlling parameter.

The rest of paper is structured as follows: section 2 introduces the single-fluid MHD model and a typical force-free paramagnetic pinch equilibrium for KTX parameters [38-40]; section 3 analyzes the linear MHD instability and its dependence on the plasma $\beta$; section 4 reports on the nonlinear simulations that demonstrate the 3D SHAx state and the favorably $\beta$ regime for the formation and maintenance of the QSH state. Finally, section 5 provides a discussion and summary.
\section{Simulation model and equilibrium}
\subsection{Simulation model}
~~~~~~~The simulations presented here are carried out using the resistive-viscous MHD model implemented in the NIMROD code [41]. The single-fluid equations can be expressed as follows:
 \begin{eqnarray}\label{eq1}
 ~~\frac {\partial{N}} {\partial{t}}+ \nabla \cdot {(N \textbf{v})} = \nabla \cdot (D\nabla N) \\
 ~\rho(\frac{\partial}{\partial{t}}+\textbf{v}\cdot \nabla)\textbf{v}=\textbf{J} \times \textbf{B}-\nabla p+\rho \nu \nabla^{2} \textbf{v}\\
 ~\frac{{N}}{(\gamma-1)}(\frac{\partial}{\partial{t}}+\textbf{v}\cdot \nabla)T=-p\nabla\cdot\textbf{v}-\nabla\cdot\textbf{q} \\
 ~\frac{\partial \textbf{B}}{\partial{t}}=-\nabla \times \textbf{E} \\
 \textbf{E}=-\textbf{v} \times \textbf{B}+\eta \textbf{J}  \\
 ~\nabla \times \textbf{B}=\mu_{0} \textbf{J} \\
 ~\nabla \cdot \textbf{B}=0\\
 ~\textbf{q}=-n[\chi_{\parallel}\hat{\textbf{b}}\hat{\textbf{b}}+\chi_{\perp}(\textbf{I}-\hat{\textbf{b}}\hat{\textbf{b}})] \cdot \nabla T
\end{eqnarray}
where $N$, $\rho$, \textbf{v}, \textbf{J}, \textbf{B}, \textbf{E}, $T$  and $p$ are the plasma number density, mass density, velocity, current density, magnetic field, electric field, electron temperature and pressure, respectively. $\gamma$, $\textbf{q}$, and $\chi_{\parallel}$ and $\chi_{\perp}$ represent the specific heat ratio, heat flux, and parallel and perpendicular thermal conductivities, respectively. $D$ is the particle diffusivity, $\eta$ the resistivity, and $\nu$ the kinematic viscosity. $\hat{\textbf{b}} \equiv \textbf{B}/{\left|\textbf{B}\right|}$ is the local magnetic direction vector.

The MHD model equations (1)-(8) involve several dimensionless parameters. The Lundquist number $S$ is defined as the ratio of the resistive diffusion timescale $\tau_{R}$ to the Alfven timescale $\tau_{A}$, $S= \tau_{R}/\tau_{A}$, where $\tau_{A} = a\sqrt{(\mu_{0}\rho)}/B$, $\tau_{R} = \mu_{0}a^2/\eta$, and $B$ represents the magnetic field strength. The magnetic Prandtl number $P_r$ is defined as the ratio of resistive and viscous times, $P_{r}$=$\tau_{R}$/$\tau_{\nu}=\mu_{0}\nu/\eta$, where $\tau_{\nu}=a^{2}/{\nu}$. The Hartmann number $H ={a V_{A}}/ {\sqrt{{\eta_{m}}{\nu}}}$. The ratio of the viscous to thermal conduction times is defined as $\tau_{\nu}/\tau_{\chi}=\chi_{\parallel}/\nu$. The plasma $\beta$ is defined as the ratio of heat and magnetic pressures, $\beta$=2$\mu_{0}$ $p$/${B}^{2}$. In addition, another parameter is the ratio of viscous to density diffusion times, given by $\tau_{\nu}/\tau_{D}=D/\nu$.

\subsection{Initial equilibrium and simulation setup}
~~~~The plasma current in the KTX experiment is currently up to $500$ $kA$, with a maximum discharge length of $100$ $ms$ [42]. The magnetic field on the axis is $B_0=0.3$ $T$, and the key equilibrium parameters are summarized in Table 1. We consider the finite $\beta$ regime in our computations, in contrast to the zero $\beta$ regime considered in the SpeCyl simulations [24-27]. The initial condition for the computations is based on a force-free paramagnetic pinch. The dimensionless parallel current density, $a\lambda=a \mu_0 \textbf{J}\cdot\textbf{B}/|\textbf{B}|^{2}$, has an on-axis value of $a\lambda(0)=3.8$ in all our computations [43]. The $q(r)$ profiles are shown in figure 1 for such a paramagnetic pinch. Both $\eta$ and $\nu$ share the same normalized radial profile given by ${\eta(r)}/{\eta_0}={\nu(r)}/{\nu_0}$=${[1+(r/a)^{20}]}^2$.

We model the KTX-RFP configuration as a periodic cylinder with the minor radius $a=0.4$ $m$, the axial length $L=2{\pi}{R_0}$, and the major radius $R_0=1.4$ $m$. We use $32\times32$ biquintic finite elements in the ($r$, $\theta$) plane and Fourier representation in the axial (z) direction. The computations involve 22 harmonics, where $ 0\leq n \leq21$, representing the axial (toroidal) harmonic number. We apply periodic boundary conditions in the axial direction and ideal wall boundary conditions in the radial direction.

\section{Linear MHD stability and $\beta$ dependence}
~~~~For the linear instability, the radial profiles of the normalized velocity perturbation $v_r$ (up) and magnetic perturbation $b_r$ along the mid-plane are shown in figure 2. It is clearly evident that the instability is located in plasma core, and the mode with $n=7$ is the most unstable for this force-free paramagnetic pinch equilibrium. The vertical dotted lines represent the radii of the resonant layers for the $m=1$, $n=7-11$ modes on the $q(r)$ profile. About each resonant surface, the radial profiles of $b_r$ are symmetric, and the radial profiles of $v_r$ are anti-symmetric, which are all of the tearing mode parity except for the $n=7$ mode, whose resonant surface is too close to the magnetic axis for its $v_r$ profile to exhibit anti-symmetry. Contrary to tokamak's results, here the locations of the resonance surfaces are all inward of the $b_r$ profile peaks [44].

As Lundquist number $S$ increases, the growth rate of the $n=7$ mode decreases slightly, whereas the growth rate of the $n=8$ mode scales as $\gamma$$\tau_A$$\propto$ $S^{-0.6}$ (figure 3), which agrees with that of the resistive tearing mode in the small $\Delta^{'}$ regime [45]. As the toroidal mode number $n$ and $\beta$ increase, the growth rate of the instability decreases, especially for $\beta<3\%$. When $3\%<\beta<10\%$, the influence of $\beta$ on the growth rate is reduced, particularly for higher $n$ modes (figure 4).

\section{SHAx state and $\beta$ dependence}
~~~~The following are the setup of the main physical parameters for the nonlinear computations based on force-free paramagnetic pinch: $B_0=0.3$ $T$, $n_0=8 \times 10^{18}$ $m^{-3}$, $\eta_0/\mu_0=23.9 $ $m^{2}/s$, $\nu=23.9 $ $m^{2}/s$, $\chi_{\parallel}=2.39$ $m^{2}/s$, $\chi_{\perp}=2.39\times 10^{-6}$ $m^{2}/s$. The nonlinear evolutions of several key RFP characteristics are displayed in figure 5 for $\beta=5\%$, including the field reversal parameter $F=B_\phi(a)/<B_\phi(r)>$ and the pinch parameter $\Theta=B_\theta(a)/<B_\phi(r)>$, where $<B_\phi(r)>$ is the volume average of the toroidal field. The computation starts from the perturbed paramagnetic pinch state and evolves into the formation of a QSH state with the dominant helicity of $m=1$, $n=7$, which experiences the fastest growth in the plasma core. The initial relaxation event disrupts the QSH state when other tearing modes reach appreciable amplitude during the period $t \leq t_1$, when the pinch parameter $\Theta$ also increases. Soon after the secondary mode reaches saturation, $\Theta$ also reaches its maximum (see figure 5, vertical dotted line $t_1$). Subsequently, along with the decay of the $m=1$ tearing modes, the magnetic energy in the edge resonant $m=0$ modes increases, and both $F$ and $\Theta$ also fall off. Each such relaxation event typically ends (see figure 5, vertical dotted line $t_2$) with a peaking in the magnetic energy of the edge resonant $m=0$ modes, and a decrease in the energies of the $m=1$ tearing modes. After that, the SHAx state recovers.

The emergence and recovery of SHAx can be also demonstrated in the time history of the Poincar¨¦ plot in the poloidal plane (figure 6). Here at $t=0.0538$ ms the initial DAx state appears, at $t=0.0745$ ms, the secondary modes grow to their peak amplitudes, resulting in magnetic field stochasticity in the core plasma region. Later at $t=0.0933$ ms, $F$ drops to a minimum and $m=0$ edge resonant modes reaches their maximum, causing magnetic field stochasticity at plasma edge. At $t=0.22$ ms, the plasma evolves to the DAx state. The plasma self-organizes to the SHAx state at $t=0.37$ ms, and a clearly shaped magnetic island appears in the plasma core with the width of helical region near $20$ cm. The SHAx lasts for about $0.4$ ms before the magnetic island becomes significantly smaller with the growth the secondary modes. Finally after $t>0.76$ ms, the island disappears, and the plasma transitions from the QSH to the MH state. For the QSH state, the shape and width of magnetic island are determined by the ratio of dominated and secondary modes magnetic energy. When the $F$ parameter reaches a minimum or the $m=0$ edge resonant modes reach their maximums at $t = 0.0933$ ms, the safety factor changes considerately in the core plasma region (figure 7).

There is an optimal range of $\beta$ that is beneficial for the formation and maintenance of the QSH state. When $\beta<4\%$, the plasma relaxes to an MH state; when $4\%\leq\beta\leq8\%$, the plasma first transitions from a DAx state to the SHAx state, and eventually return to the DAx state; when $\beta>8\%$, the plasma enters the MH state (figure 8). The peak value of dominated mode gradually decreases with the increase of $\beta$, which is consistent with the linear result. The QSH state is disrupted by relaxation events that are followed by transition into the laminar reversed field state. The frequencies of such relaxation events are higher for the $\beta$=0 regime, and the time interval between adjacent relaxation events increases with $\beta$ (figure 8(b) and (c)). The persistence time of the initial QSH state slightly increases with an increase in $\beta$ (figure 9 upper panel). For the second QSH state reappearing after the relaxation event, its duration is zero when $\beta$ is less than $4\%$. When $\beta$ is between $4\%$ and $5\%$, the second QSH state transitions from the DAx state to the SHAx state, and the duration increases. When $\beta$ is between $5\%$ and $8\%$, the duration of the QSH state gradually decreases which returns from the SHAx state to the DAx state. When $\beta$ is between $9\%$ and $10\%$, the QSH state disappears, and the plasma returns to the MH state (figure 9 lower panel).

\section{Summary and discussion}
~~~The spontaneous formation of the QSH states from a perturbed force-free paramagnetic pinch has been demonstrated in the NIMROD simulations. The dominant helicity of the initial QSH state develops from the linear mode with the maximum growth rate. Whether the plasma self-organizes into the QSH state again after a relaxation event disrupting the initial QSH state depends on the equilibrium $\beta$, and an optimal range of $\beta$ that is beneficial to the QSH formation and maintenance has been found in our simulations.

The effects of two-fluid dynamics in lower plasma current regimes on the QSH state remains less well known. It is also worth investigating whether the secondary modes become more stable during the formation of the QSH state. Additionally, other parameters that may contribute to the formation and maintenance of the QSH state should be explored in future work.

\section{Acknowledgment}
~~~We are grateful for the support of the NIMROD and the KTX teams. Author Ping Zhu would like to thank Professor C.R. Sovinec for helpful  discussions and suggestions. This study was supported by the Scientific Research Foundation of Hunan University of Arts and Science (Grant No.19BSQD37), the National Key Research and Development Program of China (Grant No. 2019YFE03050004), National Natural Science Foundation of China (Grant No. 51821005), and the U.S. Department of Energy (Grant No. DE-FG02-86ER53218). The computing work in this paper was supported by the Public Service Platform of High Performance Computing by Network and Computing Center of HUST and the Supercomputing Center of USTC.

\section{Reference}
\numrefs{1}
\item L. Marrelli, {\it et al} 2021 The reversed field pinch. {\it Nuclear Fusion} {\bf 61} 023001
\item J. Boguski, {\it et al} 2021 Direct measurements of the 3D plasma velocity in single-helical-axis RFP plasmas. {\it Physics of Plasmas} {\bf 28} 012510
\item I. H. Hutchinson, {\it et al} 1984 The structure of magnetic fluctuations in the HBTX-1A reversed field pinch.{\it Nuclear Fusion} {\bf 24} 59
\item D. F. Escande, {\it et al} 2000 Quasi-single-helicity reversed-field-pinch plasmas. {\it Physical Review Letters} {\bf 85} 1662
\item P. Martin, {\it et al} 2000 Quasi-single helicity states in the reversed field pinch: beyond the standard paradigm. {\it Physics of Plasmas} {\bf 7} 1984
\item L. Marrelli, {\it et al} 2002 Quasi-single helicity spectra in the Madison Symmetric Torus. {\it Physics of Plasmas} {\bf 9} 2868
\item L. Frassinetti, {\it et al} 2007 Spontaneous quasi single helicity regimes in EXTRAP T2R reversed-field pinch. {\it Physics of Plasmas} {\bf 14} 112510
\item R. Ikezoe, {\it et al} 2009 Quasi-periodic growth of a single helical instability in a low-aspect ratio RFP. {\it Plasma and Fusion Research} {\bf 3} 029
\item D. F. Escande, {\it et al} 2000 Single helicity: a new paradigm for the reversed field pinch.{\it Plasma Physics and Controlled Fusion} {\bf 42} B243
\item D. F. Escande, {\it et al} 2000 Chaos healing by separatrix disappearance and quasisingle helicity states of the reversed field pinch. {\it Physical Review Letters} {\bf 85} 3169
\item E. Martines, {\it et al} 2011 Equilibrium reconstruction for single helical axis reversed field pinch plasmas. {\it Plasma Physics and Controlled Fusion} {\bf 53} 035015
\item R. Lorenzini, {\it et al} 2008 Single-helical-axis states in reversed-field-pinch plasmas. {\it Physical Review Letters} {\bf 101} 025005
\item M. Gobbin, {\it et al} 2011 Vanishing magnetic shear and electron transport barriers in the RFX-mod reversed field pinch. {\it Physical Review Letters} {\bf 106} 025001
\item L. Carraro, {\it et al} 2009 Improved confinement with internal electron transport barriers in RFX-mod {\it Nuclear Fusion} {\bf 49} 055009
\item P. Piovesan, {\it et al} 2009 Magnetic order and confinement improvement in high-current regimes of RFX-mod with MHD feedback control {\it Nuclear Fusion} {\bf 49} 085036
\item R. Lorenzini, {\it et al} 2009 Self-organized helical equilibria as a new paradigm for ohmically heated fusion plasmas. {\it Nature Physics} {\bf 5} 570
\item P. Paccagnella, {\it et al} 2006 Active-feedback control of the magnetic boundary for magnetohydrodynamic stabilization of a fusion plasma. {\it Physical Review Letters} {\bf 97} 075001
\item L. Marrelli, {\it et al} 2007 Magnetic self organization, MHD active control and confinement in RFX-mod. {\it Plasma Physics and Controlled Fusion} {\bf 49} B359
\item D. Bonfiglio, {\it et al} 2013 Experimental-like helical self-organization in reversed-field pinch modeling. {\it Physical Review Letters} {\bf 111} 085002
\item M. Veranda, {\it et al} 2017 Magnetohydrodynamics modelling successfully predicts new helical states in reversed-field pinch fusion plasmas. {\it Nuclear Fusion} {\bf 57} 116029
\item E. J. Caramana, {\it et al} 1983 Nonlinear, single-helicity magnetic reconnection in the reversed-field pinch. {\it Physics of Fluids} {\bf 26} 1305
\item J. M. Finn, {\it et al} 1992 Single and multiple helicity ohmic states in reversed-field pinches. {\it Physics of Fluids B Plasma Physics} {\bf 4} 1262
\item Montgomery D., 1993 Hartmann, lundquist, and reynolds: the role of dimensionless numbers in nonlinear magnetofluid behavior. {\it Plasma Physics and Controlled Fusion} {\bf 35} B105
\item S. Cappello, {\it et al} 1992 Nonlinear plasma evolution and sustainment in the reversed field pinch {\it Physics of Fluids B Plasma Physics} {\bf 4} 1262
\item S. Cappello, {\it et al} 1996 Reconnection processes and scaling laws in reversed field pinch magnetohydrodynamics. {\it Nuclear Fusion} {\bf 4} 611
\item S. Cappello, {\it et al} 2000 Bifurcation in viscoresistive MHD: the hartmann number and the reversed field pinch. {\it Physical Review Letter} {\bf 85} 3838
\item S. Cappello, {\it et al} 2004  Bifurcation in the MHD behaviour of a self-organizing system: the reversed field pinch (RFP). {\it Plasma Physics and Controlled Fusion} {\bf 46} B313
\item D. Bonfiglio, {\it et al} 2005 Dominant electrostatic nature of the reversed field pinch dynamo. {\it Physical Review Letter} {\bf 94} 145001
\item M. Veranda, {\it et al} 2013 Impact of helical boundary conditions on nonlinear 3D magnetohydrodynamic simulations of reversed-field pinch. {\it Plasma Physics and Controlled Fusion} {\bf 55} 074015
\item D. Bonfiglio, {\it et al} 2015 Helical self-organization in 3D MHD modelling of fusion plasmas. {\it Plasma Physics and Controlled Fusion} {\bf 57} 044001
\item J. Liu, {\it et al} 2021 Three-dimensional characteristics of the quasi-single helical state in the KTX. {\it Nuclear Fusion} {\bf 61} 016017
\item Y. Zu, {\it et al} 2022 Magnetohydrodynamic dynamo effect from electrostatic drift velocity field on sustainment of reversed field pinch plasmas in three-dimensional KTX equilibrium. {\it Plasma Physics and Controlled Fusion} {\bf 64} 065002
\item K. Liu, {\it et al} 2024 Effects of magnetic helicity on 3D equilibria and self-organized states in KTX reversed field pinch. {\it Nuclear Fusion} {\bf 64} 056037
\item B. Luo, {\it et al} 2018 Resistive MHD modelling of the quasi-single helicity state in the KTX regimes. {\it Nuclear Fusion} {\bf 58} 016049
\item M. Onofri, {\it et al} 2008 Compressibility effects in the dynamics of the reversed-field pinch. {\it Physical Review Letter} {\bf 101} 255002
\item S. C. Prager, {\it et al} 1985 Driven, resistive, force-free plasmas and reversed-field-pinch physics revisited. {\it Physics of Fluids,} {\bf 28} 1155
\item J. P. Sauppe, 2015 Extended magnetohydrodynamic modelling of plasma relaxation dynamics in the reversed-field pinch. {\it University of Wisconsin-Madison. Phd thesis}
\item W. Liu, {\it et al} 2014 Progress of the Keda Torus eXperiment project in China: design and mission. {\it Plasma Physics and Controlled Fusion} {\bf 56} 094009
\item W. Liu, {\it et al} 2017 Overview of Keda Torus eXperiment initial results. {\it Nuclear Fusion} {\bf 57} 116038
\item W. Liu, {\it et al} 2019 An overview of diagnostic upgrade and experimental progress in the KTX. {\it Nuclear Fusion} {\bf 59} 112013
\item C. R. Sovinec, {\it et al} 2004 Nonlinear magnetohydrodynamics simulation using high-order finite elements. {\it Journal of Computational Physics} {\bf 195} 355
\item Z. Tao, {\it et al} 2023 Real-time MHD feedback control system in Keda Torus eXperiment. {\it Fusion Engineering and Design}  {\bf 195} 113968
\item J. P. Sauppe, {\it et al}  2017 Extended MHD modeling of tearing-driven magnetic relaxation. {\it Physics of Plasmas} {\bf 24} 056107
\item H. P. Furth, {\it et al} 1973 Tearing mode in the cylindrical tokamak. {\it Physics of Fluids} {\bf 16} 1054
\item D. C. Robinson, {\it et al} 1978 Tearing-mode-stable diffuse-pinch configurations. {\it Nuclear Fusion} {\bf 18} 939
\endnumrefs


\newpage
\begin{table}
  \centering
  \centerline{Table 1: The key equilibrium parameters used in our computations.}
  \footnotesize\rm
\begin{tabular*}{\textwidth}{@{}l*{15}{@{\extracolsep{0pt plus12pt}}l}}
\br
  ~~~Descriptions        &~~symbol    &~~Value  &~~Unit \\
\br
  ~~~Major radius  &~~~$R_0$  &~~~1.4  &~~~m \\
  ~~~Minor radius  &~~~$a$     &~~~0.4  &~~~m \\
  ~~~Toroidal magnetic field on axis &~~~$B_0$  &~~~0.3  &~~~T \\
  ~~~Number density   &~~~$n_0$  &~~~$8 \times 10^{18}$  &~~$m^{-3}$ \\
  ~~~Parallel current density on axis  &~~$a{\lambda_0}$  &~~~3.8 &~~~1\\
  ~~~Equilibrium velocity &~~~$V_0$  &~~~0 &~~~m/s\\
  ~~~$\beta$ at magnetic axis    &~~~$\beta$  &~~~$0\sim10\%$ &~~~1\\
  ~~~Resistive on axis &~~~$\eta_0$  &~~~$3.0\times10^{-5}$  &~~~$\Omega \cdot m$\\
  ~~~Kinematic viscosity on the axis &~~~$\nu_0$  &~~~$23.9$ &~~~$m/s^{2}$ \\
  ~~~Initial pinch parameter &~~~$\Theta(0)$   & ~~~1.569 &~~~1 \\
  ~~~Initial reversal parameter &~~~$F(0)$   & ~~~0.02624  &~~~1\\
  ~~~parallel thermal conduction coefficient &~~~$\chi_{\parallel}$  &~~~2.39  &~~~$m^{2}/s$   \\
  ~~~Perpendicular thermal conduction coefficient &~~~$\chi_{\perp}$  &~~~$2.39\times10^{-6}$   &~~~$m^{2}/s$  \\
  ~~~Lundquist number  &~~~S  &~~~$3.71\times10^{4}$   &~~~1\\
  ~~~Prandtl number  &~~~$P_r$    &~~~1   &~~~1\\
  ~~~Hartmann number &~~~$H$  &~~~$3.71\times 10^{4}$  &~~~1 \\
  ~~~The ratio of the viscous and thermal conduction times &~~~$\tau_{\nu}/\tau_{\chi}$   &~~~0.1   &~~~1 \\
  ~~~The ratio of viscous to density diffusion times &~~$\tau_{\nu}/\tau_{D}$   &~~~0.21   &~~~1\\
\br
\end{tabular*}
\end{table}

\newpage
\begin{figure}
  \centering
  \includegraphics[width=0.95\textwidth,clip]{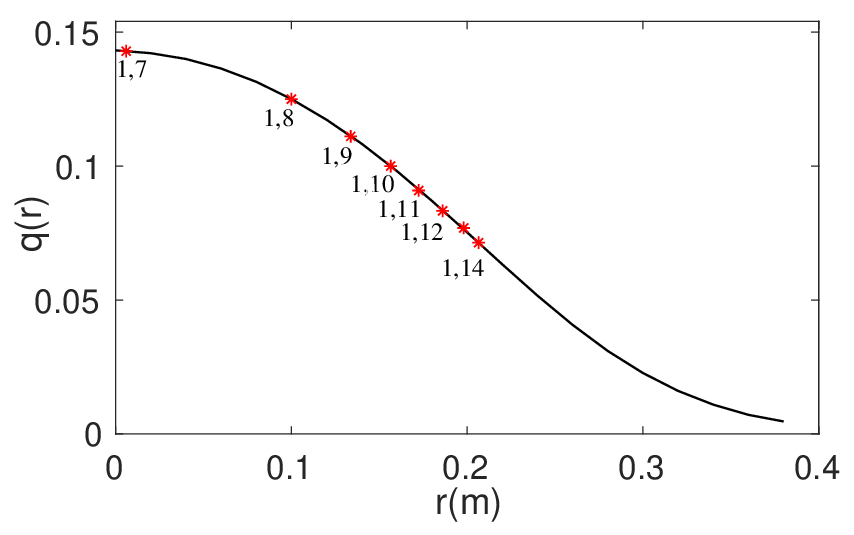}
  \caption{Safety factor as a function of the minor radius for the paramagnetic pinch with $a{\lambda(0)} = 3.8$ and $R/a=3.684$.}\label{fig.1}
\end{figure}

\newpage
\begin{figure}
  \centering
  \includegraphics[width=0.95\textwidth,clip]{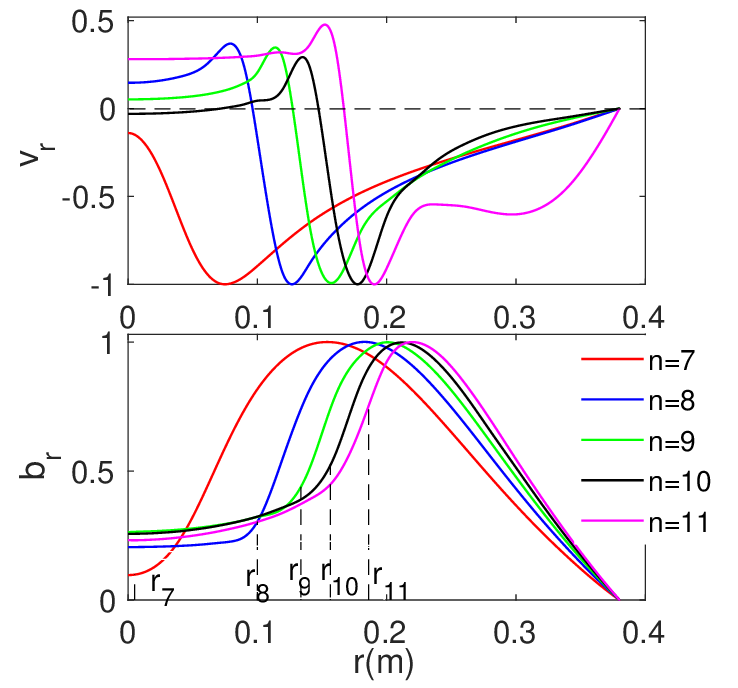}
  \caption{The radial profiles as a function of the minor radius for the normalized radial velocity perturbation $v_r$ (upper) and magnetic perturbation $b_r$ (lower) of the ($m = 1$, $n = 7-11$) modes. The vertical dotted lines represent the radii of the $m = 1$, $n = 7-11$ resonant surfaces for the given $q$ profile.}\label{fig.2}
\end{figure}

\newpage
\begin{figure}
  \centering
  \includegraphics[width=0.95\textwidth,clip]{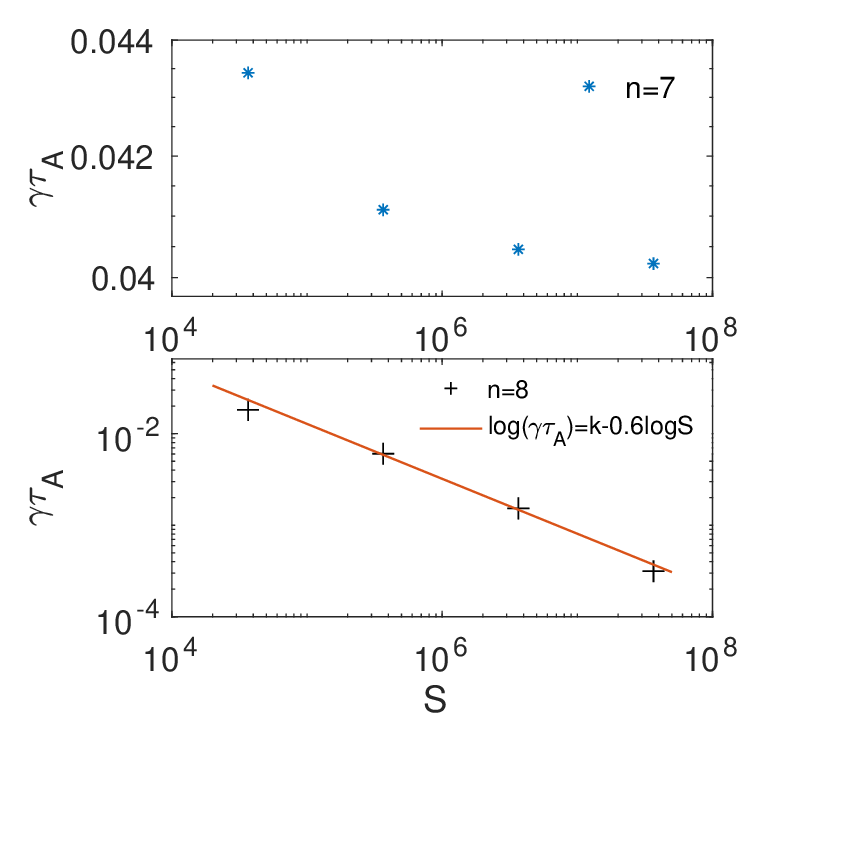}
  \caption{Linear growth rate as a function of Lundquist number $S$ on a logarithmic
scale for the $n = 7$ mode (upper) and $n=8$ mode (lower).}\label{fig.3}
\end{figure}

\newpage
\begin{figure}
  \centering
  \includegraphics[width=0.95\textwidth,clip]{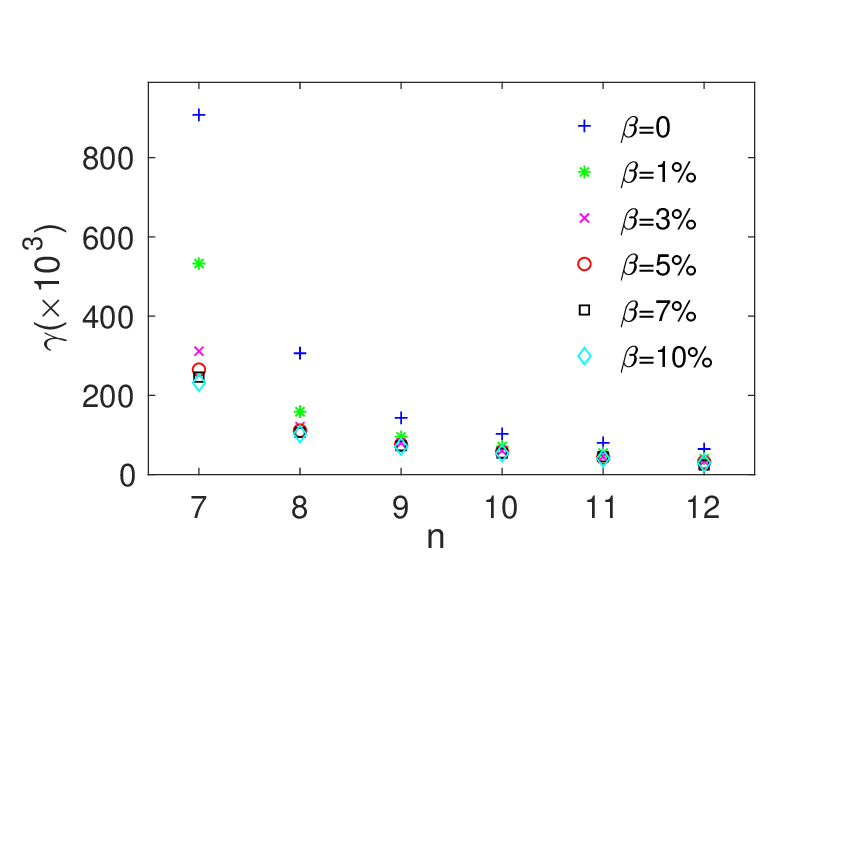}
  \caption{Linear growth rate as a function of the toroidal mode number $n$ for various $\beta$ values.}\label{fig.4}
\end{figure}

\newpage
\begin{figure}
  \centering
  \includegraphics[width=0.99\textwidth,clip]{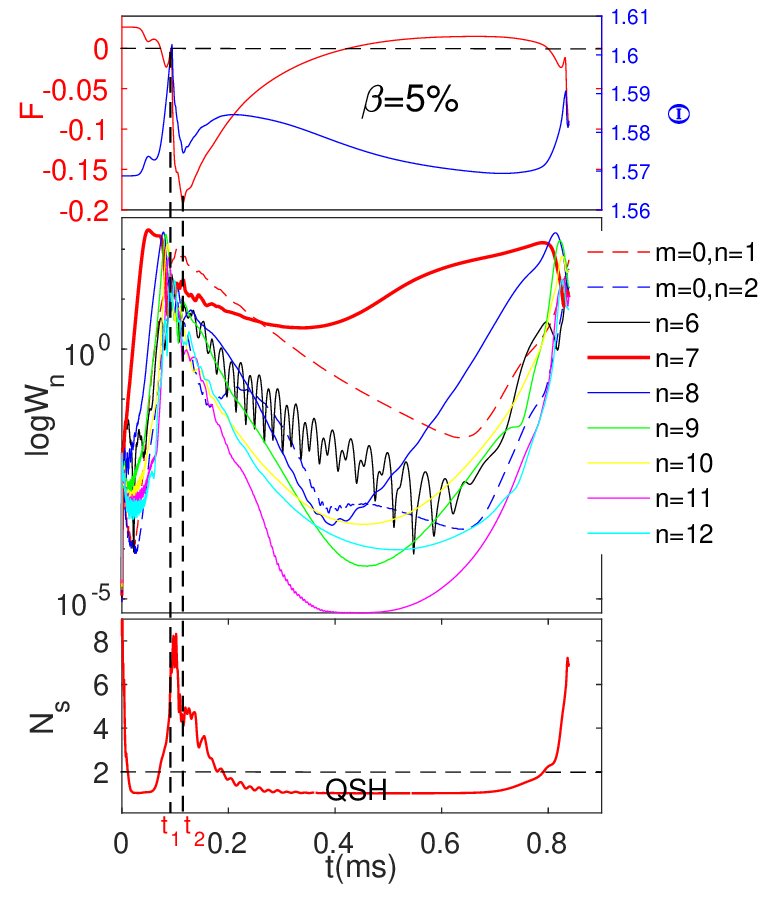}
  \caption{The upper panel shows the time evolutions of the field reversal parameter $F$ (red) and the pinch parameter $\Theta$ (blue); the middle panel shows the evolutions of the dominant magnetic fluctuation energies, including $m=0$, $n=1-2$ (dotted line) and $m=1$, $n=6-12$ (solid line) modes; and the lower panel shows the evolution of the spectral spread parameter $N_{s}=[\sum_{n=1}(W_{1,n}/\sum_{n'=1}{W_{1,n'}})^2]^{-1}$, where $W_{1,n}$ is the magnetic energy of the ($m,n$)=(1,$n$) mode. For the QSH state $N_s<2$, and $N_s=1$ is the pure SH state}.\label{fig.5}
\end{figure}

\newpage
\begin{figure}
  \centering
  \includegraphics[width=6.1cm,height=6.1cm,angle=270,clip]{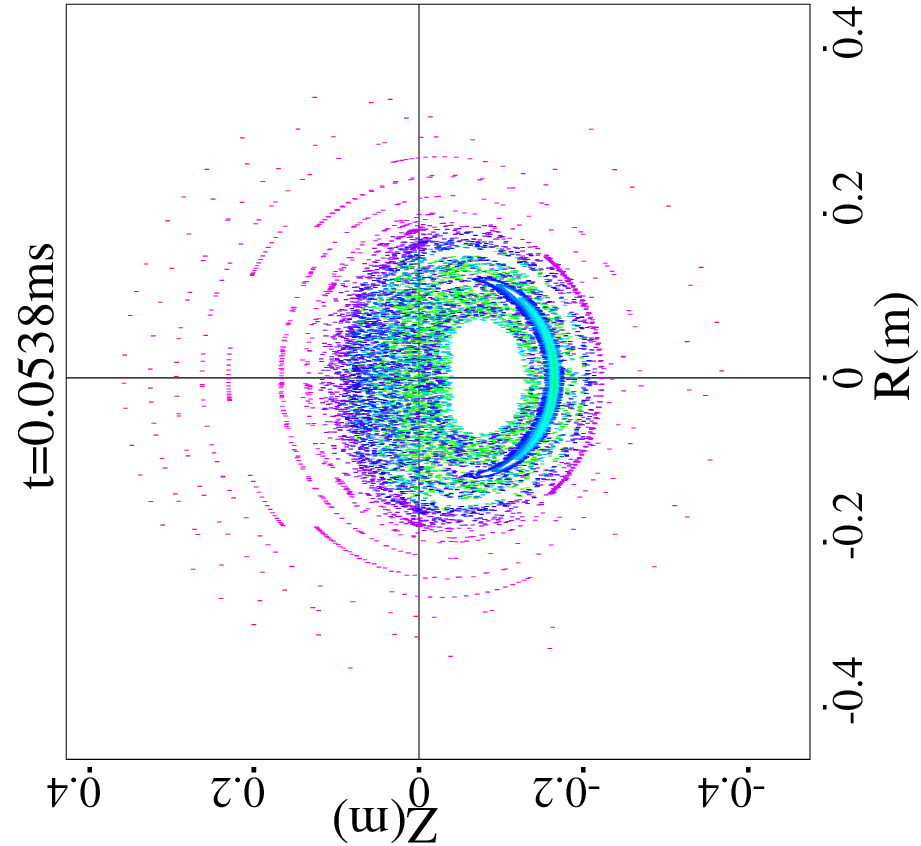}
  \includegraphics[width=6.1cm,height=6.1cm,angle=270,clip]{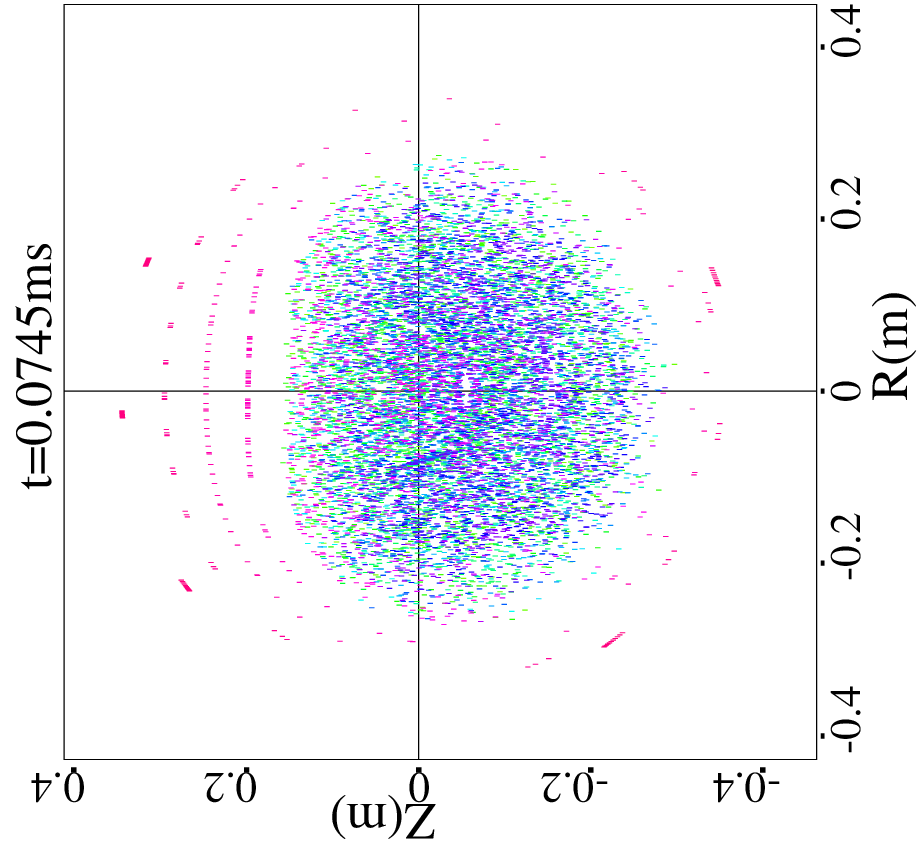}
  \includegraphics[width=6.1cm,height=6.1cm,angle=270,clip]{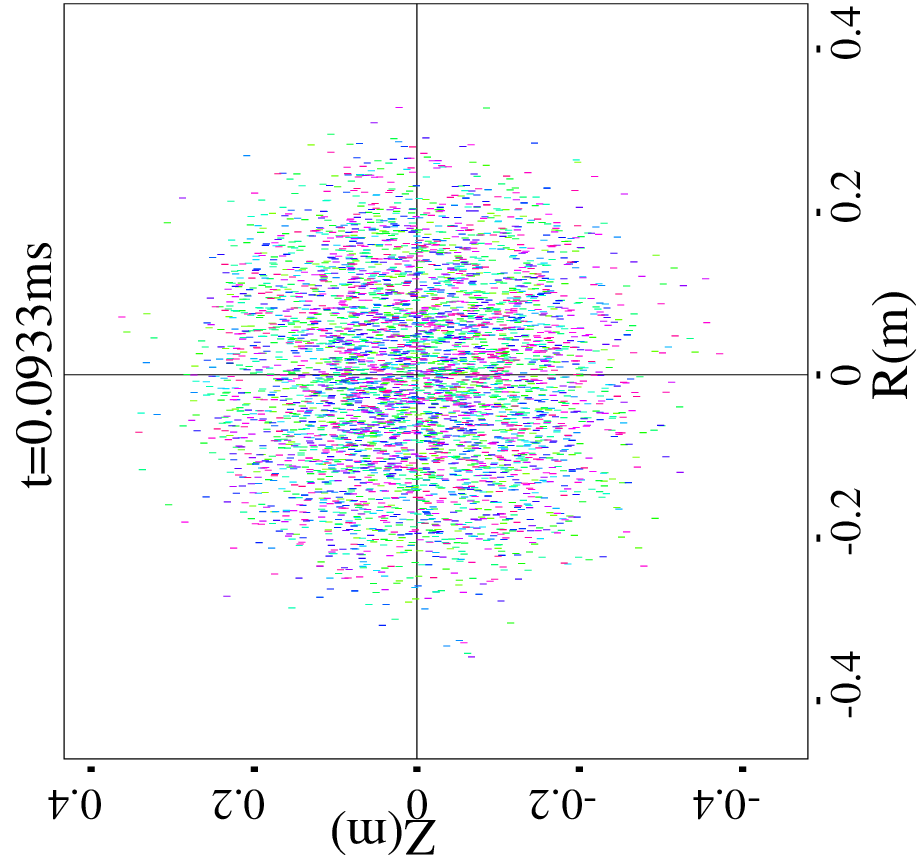}
  \includegraphics[width=6.1cm,height=6.1cm,angle=270,clip]{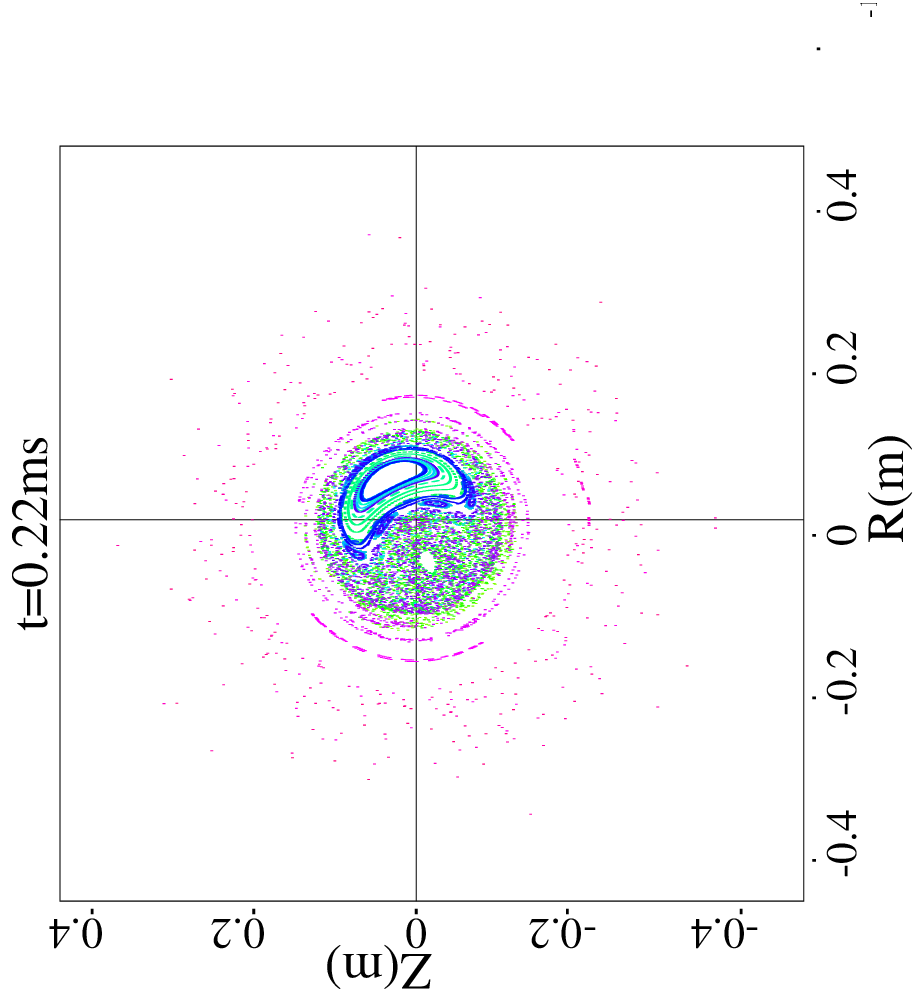}
  \includegraphics[width=6.1cm,height=6.1cm,angle=270,clip]{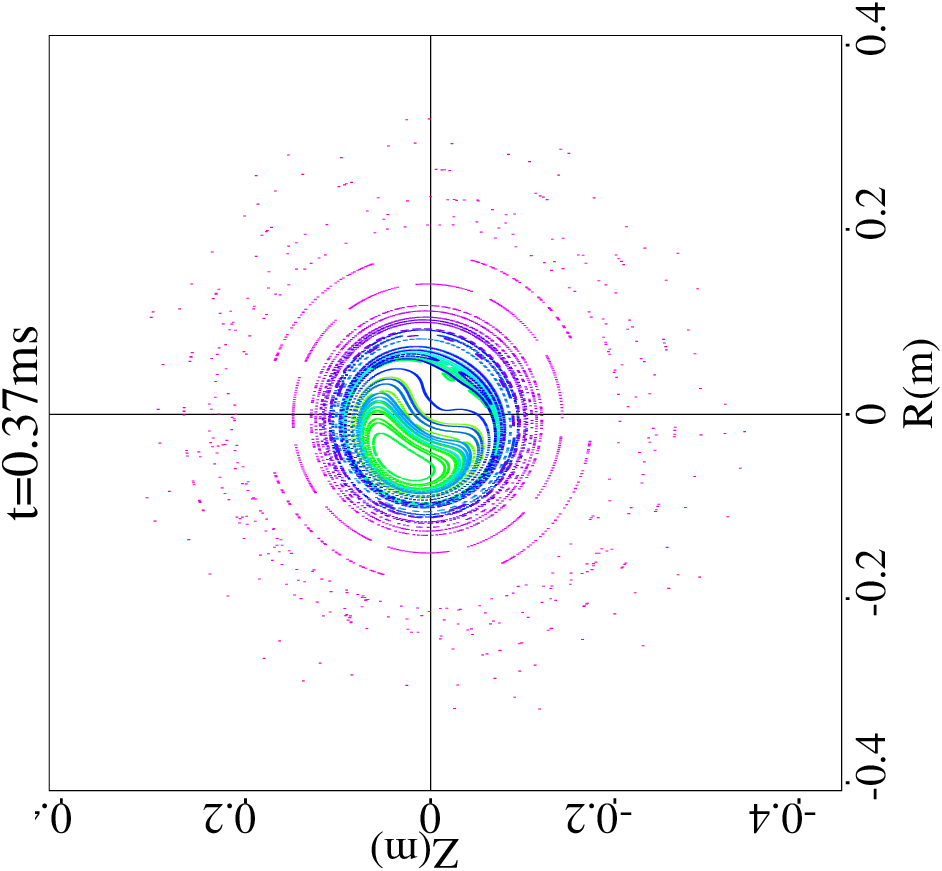}
  \includegraphics[width=6.1cm,height=6.1cm,angle=270,clip]{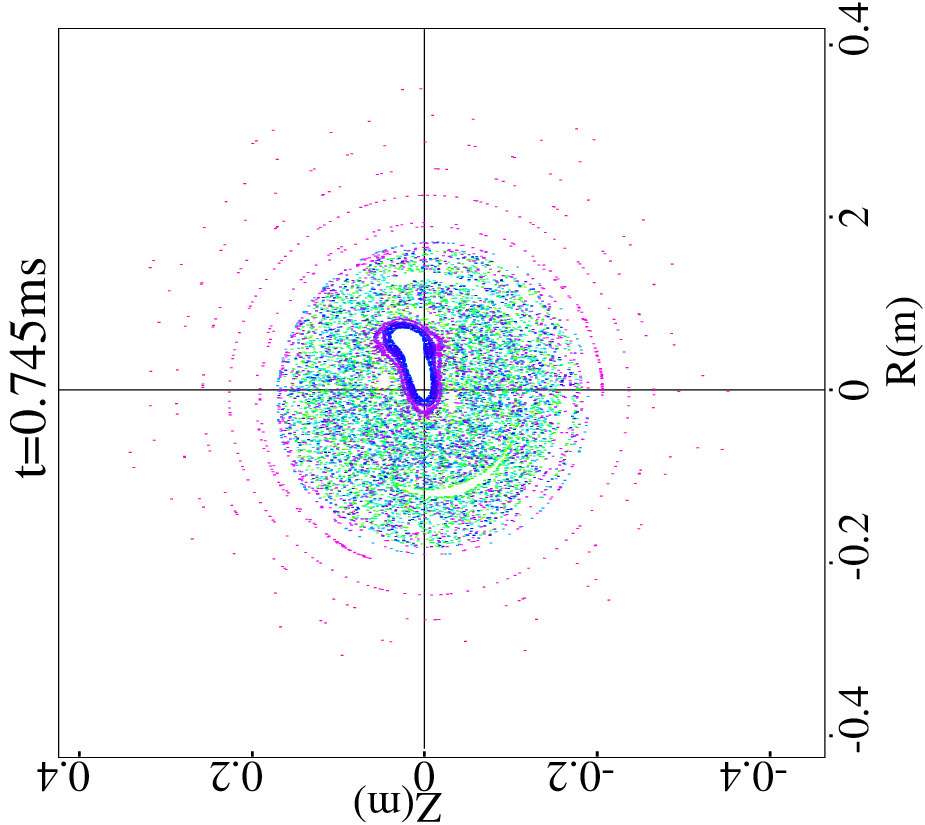}
  \caption{Snapshots of the Poincar¨¦ plots in the poloidal plane for the simulation case shown in figure 5, at time $t=$0.0538ms, 0.0745ms, 0.0933ms, 0.22ms, 0.37 ms and 0.745ms, respectively.}\label{fig.6}
\end{figure}

\newpage
\begin{figure}
  \centering
  \includegraphics[width=0.98\textwidth,clip]{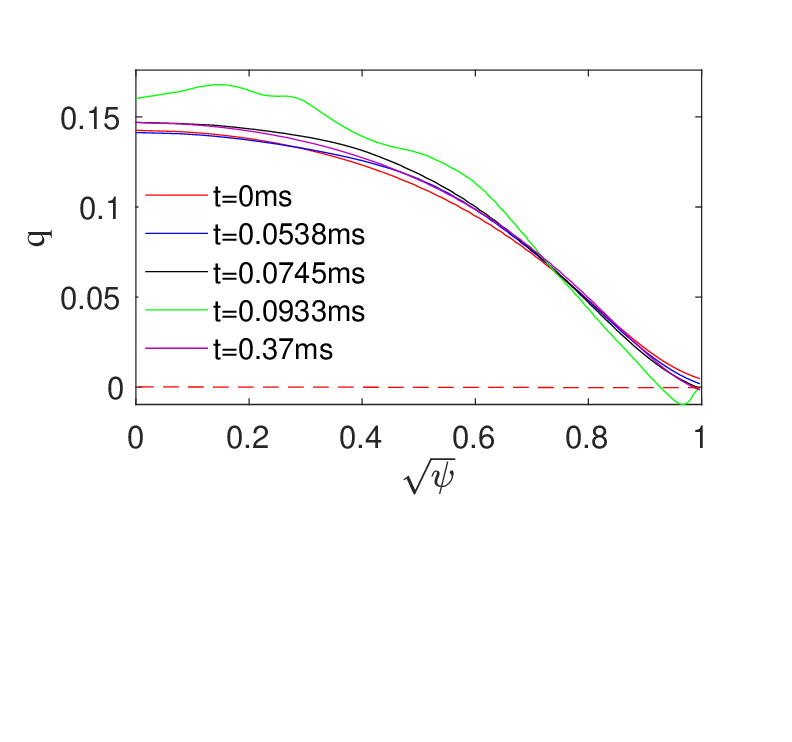}
  \caption{The safety factor as a function of $\sqrt{\psi}$ at a sequence of time slices.}\label{fig.7}
\end{figure}

\newpage
\begin{figure}
  \centering
  \includegraphics[width=7cm,height=10cm,clip]{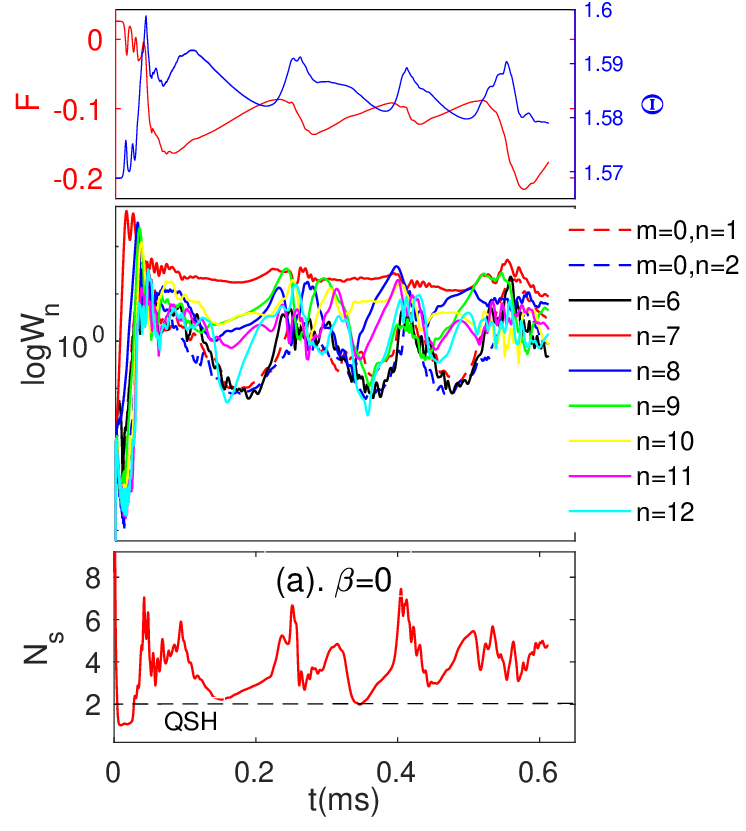}
  \includegraphics[width=7cm,height=9.9cm,clip]{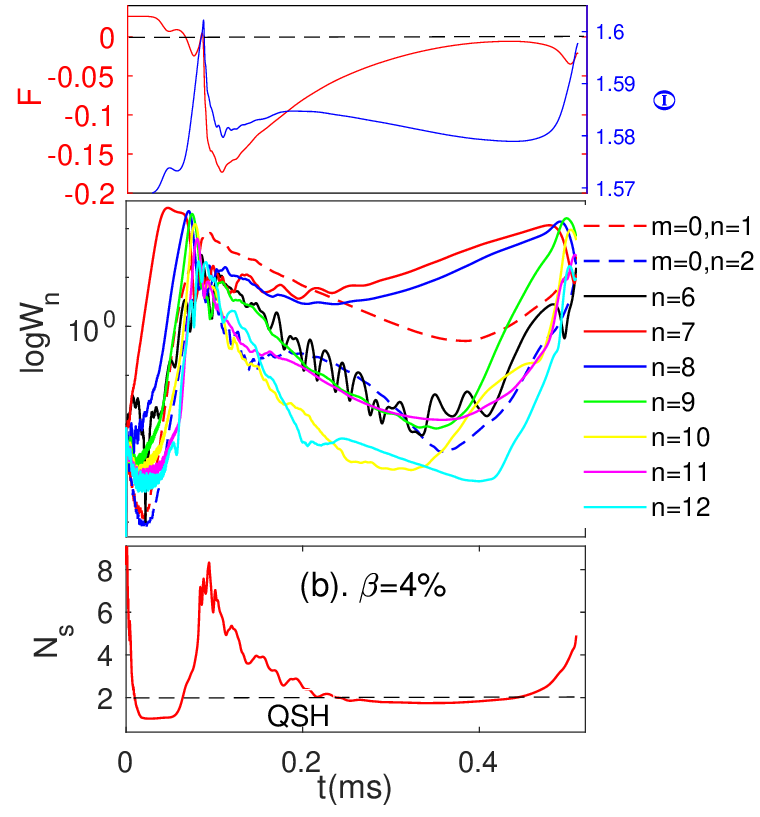}
  \includegraphics[width=7cm,height=10cm,clip]{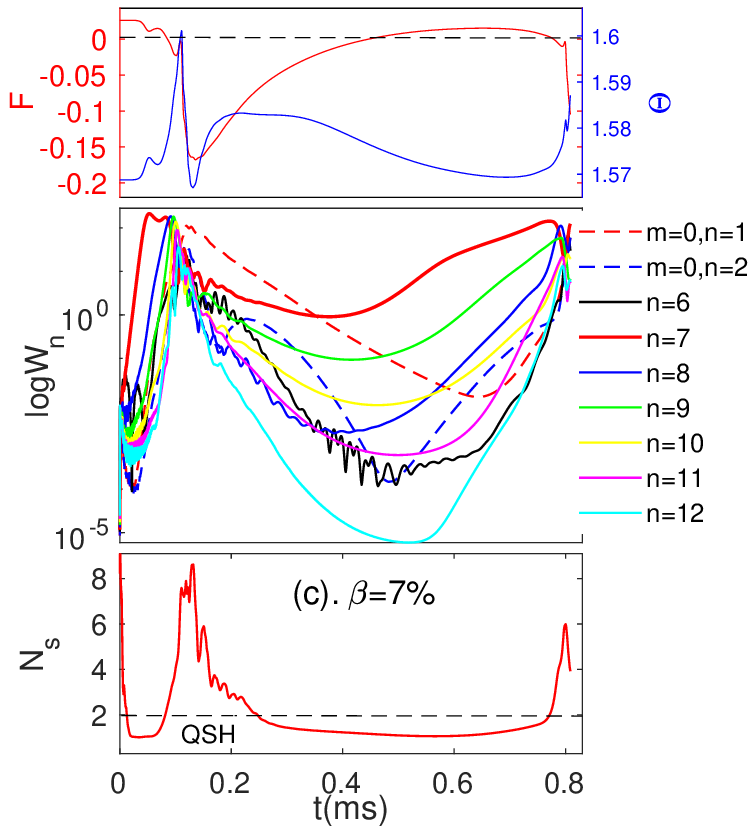}
  \includegraphics[width=7cm,height=9.9cm,clip]{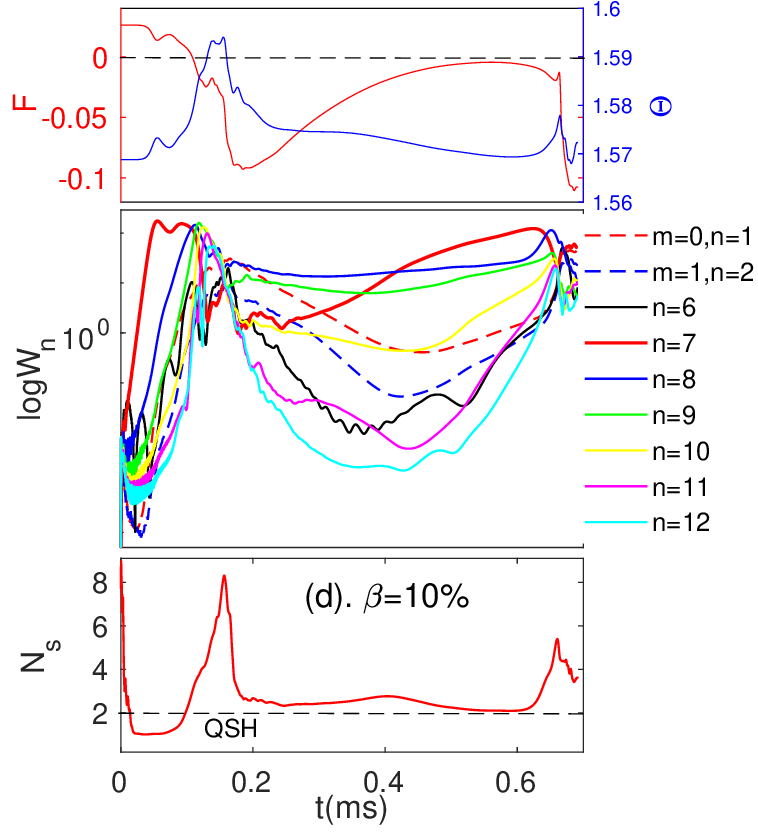}
  \caption{The field reversal $F$ and the pinch parameter $\Theta$ (upper), spectral magnetic energy (middle) and the spectral spread $N_s$ (lower) an functions of time for (a):$\beta=0$, (b):$\beta=4\%$, (c):$\beta=7\%$, (d):$\beta=10\%$.}\label{fig.8}
\end{figure}

\newpage
\begin{figure}
  \centering
  \includegraphics[width=0.99\textwidth,clip]{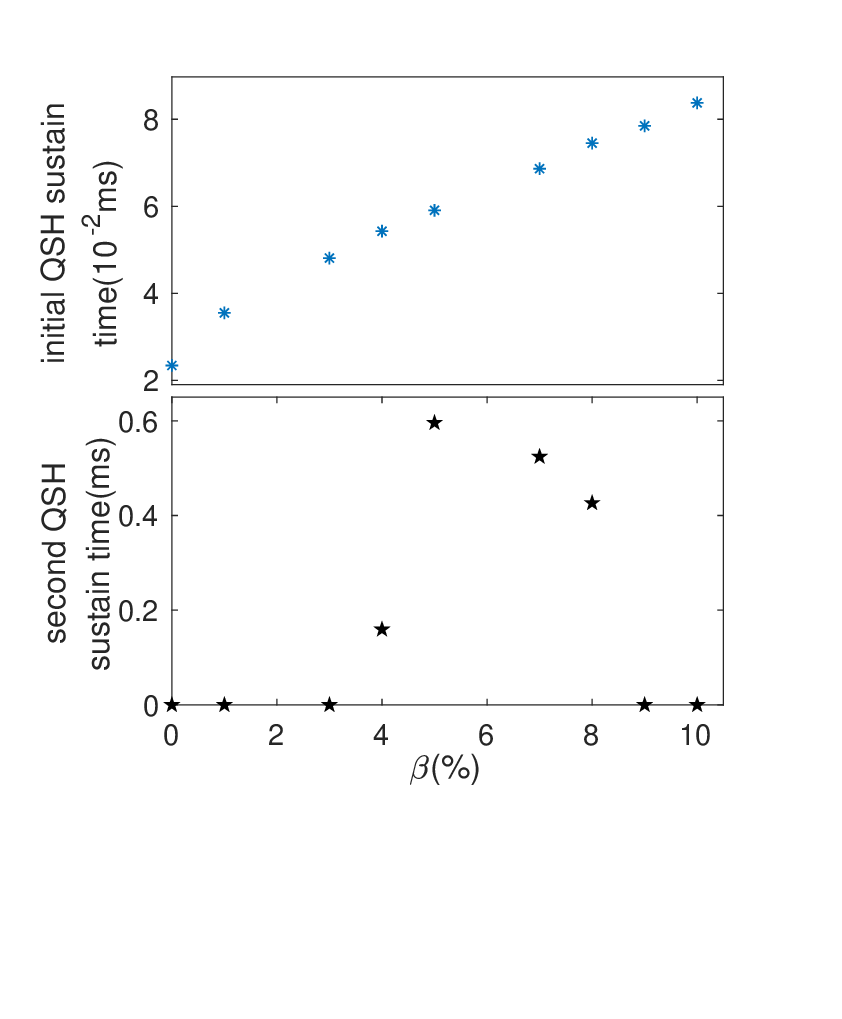}
  \caption{The first the QSH persistence (upper) and second persistence time (lower) for various $\beta$ values.}\label{fig.9}
\end{figure}
\end{document}